\documentclass[aps,prl,twocolumn,showpacs,superscriptaddress]{revtex4}
\usepackage{graphicx}
\usepackage{amsmath}
\usepackage{amssymb}
\usepackage{colordvi}
\usepackage{mathrsfs}
\usepackage{bm}
\usepackage{verbatim}
\usepackage{dcolumn}
\usepackage{bm}
\usepackage{epsfig}
\usepackage{subfigure}

\begin{document}
\title{High-Temperature Quantum Anomalous Hall Effect in $n$-$p$ Codoped Topological Insulators}
\author{Shifei Qi}
\affiliation{International Center for Quantum Design of Functional Materials, Hefei National Laboratory for Physical Sciences at Microscale, and Synergetic Innovation Center of Quantum Information and Quantum Physics, University of Science and Technology of China, Hefei, Anhui 230026, China}
\affiliation{School of Chemistry and Materials Science, Shanxi Normal University, Linfen, Shanxi 041004, China}
\author{Zhenhua Qiao}
\affiliation{International Center for Quantum Design of Functional Materials, Hefei National Laboratory for Physical Sciences at Microscale, and Synergetic Innovation Center of Quantum Information and Quantum Physics, University of Science and Technology of China, Hefei, Anhui 230026, China}
\affiliation{Department of Physics, University of Science and Technology of China, Hefei, Anhui 230026, China}
\author{Xinzhou Deng}
\affiliation{International Center for Quantum Design of Functional Materials, Hefei National Laboratory for Physical Sciences at Microscale, and Synergetic Innovation Center of Quantum Information and Quantum Physics, University of Science and Technology of China, Hefei, Anhui 230026, China}
\affiliation{Department of Physics, University of Science and Technology of China, Hefei, Anhui 230026, China}
\author{Ekin D. Cubuk}
\affiliation{School of Engineering and Applied Sciences, Harvard University, Cambridge, Massachusetts 02138, USA}
\author{Hua Chen}
\affiliation{Department of Physics, The University of Texas at Austin, Austin, Texas 78712, USA}
\author{Wenguang Zhu}
\affiliation{International Center for Quantum Design of Functional Materials, Hefei National Laboratory for Physical Sciences at Microscale, and Synergetic Innovation Center of Quantum Information and Quantum Physics, University of Science and Technology of China, Hefei, Anhui 230026, China}
\affiliation{Department of Physics, University of Science and Technology of China, Hefei, Anhui 230026, China}
\author{Efthimios Kaxiras}
\affiliation{School of Engineering and Applied Sciences, Harvard University, Cambridge, Massachusetts 02138, USA}
\author{S. B. Zhang}
\affiliation{Department of Physics, Applied Physics and Astronomy, Rensselaer Polytechnic Institute, Troy, New York 12180-3590, USA}
\author{Xiaohong Xu}
\email[Correspondence author:~~]{xuxh@dns.sxnu.edu.cn}
\affiliation{School of Chemistry and Materials Science, Shanxi Normal University, Linfen, Shanxi 041004, China}
\author{Zhenyu Zhang}
\email[Correspondence author:~~]{zhangzy@ustc.edu.cn}
\affiliation{International Center for Quantum Design of Functional Materials, Hefei National Laboratory for Physical Sciences at Microscale, and Synergetic Innovation Center of Quantum Information and Quantum Physics, University of Science and Technology of China, Hefei, Anhui 230026, China}
\date{\today{}}
\begin{abstract}
  {The quantum anomalous Hall effect (QAHE) is a fundamental quantum transport phenomenon that manifests as a quantized transverse conductance in response to a longitudinally applied
electric field in the absence of an external magnetic field, and promises to have immense application potentials in future dissipation-less quantum electronics. Here we present a novel kinetic pathway to realize the QAHE at high temperatures by $n$-$p$ codoping of three-dimensional topological insulators. We provide proof-of-principle numerical demonstration of this approach using vanadium-iodine (V-I) codoped Sb$_2$Te$_3$ and demonstrate that, strikingly, even at low concentrations of $\sim$2\% V and $\sim$1\% I, the system exhibits a quantized Hall conductance, the tell-tale hallmark of QAHE, at temperatures of at least $\sim$ 50 Kelvin, which is three orders of magnitude higher than the typical temperatures at which it has been realized so far. The proposed approach is conceptually general and may shed new light in experimental realization of high-temperature QAHE.}
\end{abstract}
\maketitle

Two-dimensional electron systems (2DESs) are versatile playgrounds for complex quantum transport phenomena that are of interest both from fundamental and application points of view. One prominent example is the quantum Hall effect (QHE) ~\cite{QHE1,QHE2}, a quantum analogue of the classical Hall effect, whereby the application of a magnetic field to a 2DES results in the quantization of the transverse conductance and a vanishing longitudinal conductance. Crucially, the mechanisms underlying this phenomenon are the formation of an insulating bulk and remarkable chiral conducting edge states (Fig. 1a). This striking unidirectionality makes backscattering impossible, and renders transport along these edge states remarkably robust against any impurities~\cite{QHE3,QHE4}. This robustness essentially renders the edges of 2DESs manifesting the QHE as perfectly conducting one-dimensional (1D) wires, which would be of great interest as potential building blocks, such as interconnects between chips, in dissipation-less electronics. However, a major constraint for practical use of this fundamental quantum transport phenomenon is the requirement of very strong magnetic fields, which is impractical for realistic applications.

Realizing the quantum anomalous Hall effect (QAHE) would circumvent this problem because it manifests the same hallmark features as the QHE, but does not require the application of an external magnetic field ~\cite{Haldane}. The reason why the QAHE does not require the application of an external field is because the latter's role is played by an intricate cooperation between the intrinsic magnetism that breaks the time-reversal symmetry and spin-orbit coupling in the 2D (quasi-2D) insulating host material. Initial proposals for realizing the QAHE were based on honeycomb lattice models ~\cite{Haldane}. Since then, especially after the experimental discoveries of graphene (with an inherent honeycomb lattice structure) ~\cite{graphene} and topological insulators (TIs) ~\cite{TopologicalInsulator1,TopologicalInsulator2}, much effort has been made to exploring new platforms for realizing QAHE in these and related systems ~\cite{proposal1,proposal2,proposal3,proposal4,proposal5,proposal6,proposal7,proposal8,proposal9,proposal10,proposal11,proposal12}. TIs are inherently superior to graphene for this purpose because of the formers' intrinsically strong spin-orbit coupling, which narrows the search for suitable materials to those that could harbor intrinsic ferromagnetism. One possible way to introduce ferromagnetism in TIs is by introducing magnetic dopants ~\cite{proposal4,TI-magnetism1,TI-magnetism2,TI-magnetism3}, similar to the approach taken in realizing diluted magnetic semiconductors~\cite{DMS}. This approach has led to the first experimentally realized QAHE systems in Cr-doped (Bi,Sb)$_2$Te$_3$ thin films ~\cite{ChangCuiZu,ExperimentalQAHE1,ExperimentalQAHE2}. Yet to date, all the experimentally observed QAHE were measured only at extremely low temperatures, which would still prohibit potential applications of this fundamental quantum transport phenomenon.

\begin{figure}
  \includegraphics[width=6.0cm,angle=0]{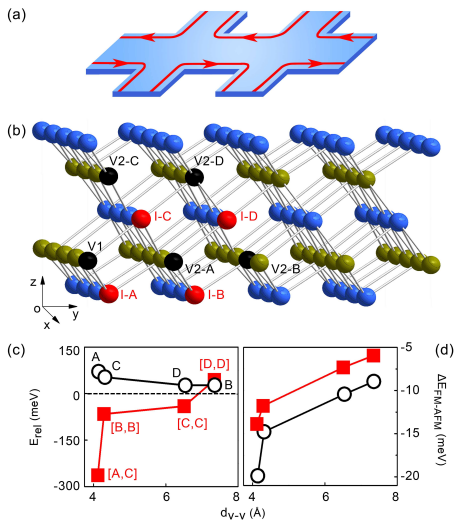}
  \caption{(a) Schematic of a Hall-bar device with chirally propagating edge modes. (b) The $4\times 4 \times 1$ supercell of Sb$_2$Te$_3$ with only one of three QLs shown for clarity. The notations are: V1 is the site, where one V atom substitutes one Sb atom; V2-A - V2-D are sites that can be occupied by a second V dopant; I-A - I-D designate sites that can be occupied by one I dopant. (c) V-V interaction energy $E_{\rm rel}$ vs. the V-V separation $d_{\rm V-V}$ of V doped (open black circles) and V-I codoped Sb$_2$Te$_3$ (filled red squares), respectively. For the V doped case, the label of each open circle indicates the position of the second V atom. In the V-I codoped case, one I is fixed at I-A, and the two labels in each square bracket denote the positions of the second V and second I, respectively. (d) Magnetic coupling between two V dopants vs. $d_{\rm V-V}$ of V doped (open black circles) and V-I codoped (filled red squares) Sb$_2$Te$_3$, respectively.}
\label{Fig1}
\end{figure}
In this Letter, we propose a versatile approach to realize high-temperature QAHE by using compensated $n$-$p$ codoping to establish high-temperature ferromagnetic order in 3D-TIs. This doping approach in large part does not alter the bulk gaps and thus addresses the primary challenge that has hindered experimental realization of high-temperature QAHE in TIs so far~\cite{ChangCuiZu,ExperimentalQAHE1,ExperimentalQAHE2}. We provide numerical proof-of-principle demonstration of this approach using a prototypical example, by codoping Sb$_2$Te$_3$ with magnetic $p$-type vanadium (V) and nonmagnetic $n$-type iodine (I) dopants. These dopants are chosen because of their preference to form $n$-$p$ pairs due to their mutual electrostatic attraction, thereby enhancing the solubility of both dopants. More importantly, we show that whereas doping with V alone would undesirably shift the Fermi level into the valence band and shrink the bulk band gap, codoping with I recovers the insulating nature while also significantly restores the intrinsic bulk band gap of the TI. Moreover, for all the codoping concentrations considered in our study, the coupling between the hard magnetic V dopants always gives rise to a stable ferromagnetic order. Strikingly, even at a low codoping concentration of 2.08\% V and 1.39\% I (which amounts to compensated one-to-one V-I codoping), an estimated Curie temperature about 49 Kelvin can be reached. Aside from the bulk properties, the band gaps associated with the surface states of the corresponding thin film structures can also be as large as 53 meV, much larger than what is needed to sustain ferromagnetic order at the high Curie temperatures. Explicit calculations of the Berry curvatures further confirm that the V-I codoped Sb$_2$Te$_3$ thin films can harbor QAHE with the Hall conductance quantized to be $\sigma_{xy}=e^2/h$ . The proposed approach thus offers a versatile route to the realization of high-temperature QAHE in different TIs and related materials.
\begin{figure}
  \includegraphics[width=7cm,angle=0]{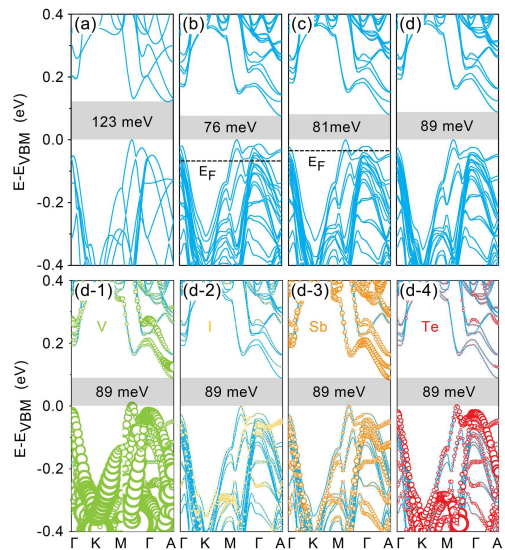}
  \caption{  Band structures of V doped and V-I codoped Sb$_2$Te$_3$. (a) Pure  Sb$_2$Te$_3$ with a bulk band gap of 123 meV; (b) For 2.08\% V doping, the Fermi level shifts into the valence band, and the bulk gap decreases to 76 meV; (c) Additional 0.69\% I codoping slightly moves the Fermi level towards the bulk gap and enlarges the gap to 81 meV; (d) A further increase of I codoping to 1.39\% compensates the V dopants and shifts the Fermi level into the bulk gap of 89 meV, effectively recovering the insulating nature of the system. (d-1)-(d-4) Different characters of the bands shown in panel (d), obtained by projecting the Kohn-Sham states onto the local orbitals of a single atom for each element.} \label{Fig2}
\end{figure}

Our first-principles calculations were performed using the projected augmented-wave method ~\cite{PAW} as implemented in the Vienna Ab-initio Simulation Package (VASP)~\cite{VASP1}. The generalized gradient approximation (GGA) of Perdew-Burke-Ernzerhof (PBE) type was used to treat the exchange-correlation interaction ~\cite{GGA}. A $4\times 4\times1$ Sb$_2$Te$_3$ supercell is chosen to study the magnetic coupling between a pair of V dopants. Each supercell is composed of three quintuple layers (QLs), including 96 Sb atoms and 144 Te atoms. For thin film calculations, the film thickness was chosen to be six QLs, including 192 Sb and 288 Te atoms. A vacuum buffer space of 30~\AA was used to prevent the coupling between adjacent slabs. The kinetic energy cutoff was set to 250 eV. During structural relaxation, all atoms were allowed to relax until the Hellmann-Feynman force on each atom is smaller than 0.01 eV/\AA. The Brillouin-zone integration was carried out by respectively using $3\times 3 \times 2$ and $1\times 1$ Monkhorst-Pack grids for bulk and thin film systems. Unless mentioned otherwise, spin-orbit coupling was considered in all calculations, and the GGA+U method was used with on-site repulsion parameter U=3.00 eV~\cite{AnotherU} and exchange parameter J =0.87 eV. The Curie temperature $T_C$ was estimated within the mean-field approximation $k_BT_C = \frac{2}{3}x \sum_{i \neq 0} J_{0i}$~ \cite{CurieTemperature}, where $k_B$ is the Boltzmann constant, $x$ is the dopant concentration, and $J_{0i}$ is the on-site exchange parameter obtained from the total energy difference between the ferromagnetic and anti-ferromagnetic configurations.

To elucidate the role of codoping in realizing high-temperature QAHE, we first investigate the energetics and magnetic behavior of Sb$_2$Te$_3$ solely doped with V, whereby V dopants substitute Sb atoms ~\cite{V-SbTe}. The interaction energy between two V atoms is defined as
\begin{eqnarray}\label{eq:erel}
E_{\rm rel}=E_{\rm 2V}+E_{\rm 0V}-2E_{\rm 1V},
\end{eqnarray}
where $E_{ i \rm V}$ is the total energy of the Sb$_2$Te$_3$ systems containing \textit{i} (\textit{i}=0, 1, 2) V atoms. We found that the V-V interaction is repulsive (Fig. 1c), with $E_{\rm rel} > 0$ for all the positions considered. Thus the energetics dictate that the V dopants arrange according to a diluted distribution. We also found that the V-V magnetic coupling is ferromagnetic for all the V-V separation considered in our study, strongly indicating a preponderance toward a diluted ferromagnetic (Fig. 1d) - the magnetic coupling between two V moments is defined as the total energy difference between the ferromagnetic and anti-ferromagnetic configurations.

One important finding from our comparison of the electronic band structures of pristine Sb$_2$Te$_3$ (Fig. 2a) and V-doped  Sb$_2$Te$_3$ (Fig. 2b) is that solely doping the TI with V-dopants drives the system from an insulator (Fig. 2a) to a $p$-type semiconductor (Fig. 2b), even at a low concentration of 2.08\%. The Fermi level is thereby shifted to the valence band (Fig. 2b), which is consistent with previous experimental findings ~\cite{V-SbTe}. V doping also decreases the bulk gap from 123 meV to 76 meV at the 2.08\% V concentration, which originates from the V contribution to the valence band maximum (VBM) (Fig. 2(d-1)) ~\cite{WGZhu}. Consequently, V doping alone cannot convert Sb$_2$Te$_3$ into a magnetic TI.

We now demonstrate that to recover the insulating nature of  Sb$_2$Te$_3$ it is necessary to compensate the extra charge introduced by V dopants, by adding an additional type of dopant with opposite charge, i.e., in this case an $n$-type dopant. For the purpose of our proof-of-principle demonstration we chose I as the dopant to substitute Te in Sb$_2$Te$_3$, which results in an $n$-type doping. This choice was inspired by the fact that SbTeI and BiTeI are naturally existing materials~\cite{SbTeI}. Moreover, I possesses larger spin-orbit coupling than Te, which is likely to have the added benefit for enhancing the topological phases~\cite{Nagaosa1,Nagaosa2}. Using an $n$-$p$ codoping approach has also the added benefits of stabilizing the constituent dopants (because of their mutual electrostatic attraction, which leads to the codopants forming closely spaced pairs during the doping process), increasing their solubility, and potentially leading to a higher magnetic transition temperature, as in the case of diluted magnetic semiconductors~\cite{npCodopedDMS}. Additionally, we note that the experimentally fabricated Sb$_2$Te$_3$ is usually a light-doped TI (therefore a conductor) due to the defects, which, based on our compensation principle, can be potentially tuned to be an insulator by properly adjusting the ratio of $n$- and $p$-doping.

Codoping Sb$_2$Te$_3$ with V and I gives rise to striking changes in the electronic band structure. First, codoping with 0.69\% I, in addition to 2.08\% V doping, shifts the Fermi level upwards (Fig. 2c). For this low I-doping concentration, the system is still a $p$-type semiconductor. When the I concentration further increases to 1.39\% (which entails one-to-one V-I codoping), the system recovers its insulating phase, and the bulk gap also increases to 89 meV (Fig. 2d). Further analysis of the characters of different elements in the electronic band structure (Figs. 2d(1)- 2d(4)) reveals that the valence band close to the Fermi level is mainly dominated by V, Sb, and Te atoms, while I contributes mainly to the lower part of the valence bands. These results directly reflect the strong bonding interaction between I and the surrounding V/Sb atoms, because I has larger electronegativity than Te, and the resulting large energy difference between the bonding and anti-bonding orbitals helps to enlarge the bulk gap. Therefore, the introduction of I not only shifts the Fermi level into the bulk gap but also enlarges the gap size, which are the salient features of the compensated $n$-$p$ codoping scheme. Taken together, these results demonstrate that it is possible to preserve an insulating state for bulk Sb$_2$Te$_3$ even in the presence of doping, provided that one-to-one compensated V-I codoping is used.

Preserving the insulating phase during doping is one key requirement for realizing high-temperature QAHE. The second requirement is to achieve a strong ferromagnetism. We will therefore examine the magnetic properties of the V-I codoped system, and in particular elucidate under what conditions ferromagnetic order is favored. Before doing so, it is important to investigate in more detail how two separate V-I pairs interact in the host system. Our calculations show, first, that the pair-pair interaction energies $E_{\rm rel}$ are attractive (Fig. 1c). A second conclusion from our calculations is that the magnetic coupling between two magnetic moments associated with V dopants is ferromagnetic, with $\Delta E_{\rm FM-AFM}<0$, for all the considered dopant separation ranges. In addition, we found that, when compared to the pure V-doped case, the resultant magnetic moment of a V-I pair is larger, but the coupling strength between two such pairs is somewhat smaller. This can be attributed to the fact that the V-I codoped system becomes insulating, whereas the V-doped system is metallic and has delocalized charge carriers that mediate stronger magnetic coupling. Taken together, these findings indicate that the $n$-$p$ codoping scheme efficiently leads to a stable ferromagnetic state even at relatively low codoping concentrations, which is highly advantageous compared to other approaches that rely on heavy magnetic doping~\cite{ChangCuiZu,ExperimentalQAHE1,ExperimentalQAHE2}.

The ferromagnetism induced using our codoping scheme is highly robust and has the potential to exist at high temperatures. For the particular doping concentrations in our proof-of-principle demonstrations, using the magnetic coupling strengths extracted from our first-principles calculations, we estimate the Curie temperature within mean-field theory~\cite{CurieTemperature} to be $T_{\rm C}=49~(152)$ Kelvin for codoping concentrations of 2.08\% V and 1.39\% I, respectively [the results were obtained within GGA+U (GGA) by placing two V-I pairs in diluted configuration at one QL]. With further increase of the codoping concentration it is possible to raise the Curie temperature, as long as the dopant pair concentrations are still low enough to be away from potential clustering of the dopants~\cite{Carbon}. Therefore, the present estimations give a lower limit on the Curie temperature at such a low codoping concentration.

Our proof-of-principle demonstration thus shows that V-I codoped  Sb$_2$Te$_3$ is a ferromagnetic insulator with a sizable gap, making the system a highly desirable candidate for realizing high-temperature QAHE. To demonstrate that such systems can indeed harbor QAHE for a wide range of codoping concentrations, we have calculated and analyzed the band structures of intrinsic and V-I codoped  Sb$_2$Te$_3$ thin films and shown that these systems yield the quantized Hall conductance $\sigma_{xy}=e^2/h$ expected for the QAHE, i.e., without the external magnetic field. To that end, first, we have shown that these codoped systems harbor large energy gaps around the Dirac point. In the absence of dopants, the surface of a Sb$_2$Te$_3$ thin film hosts a massless Dirac fermion, manifested as the trademark linear Dirac dispersion near the gamma point~\cite{Sb2Te3} (Fig. 3a). After introducing magnetic dopants, the Dirac fermion acquires a finite mass due to the presence of the intrinsic ferromagnetism, which in principle can give rise to the QAHE~\cite{TopologicalInsulator2}. Here we show the calculated band structures and Berry curvatures, along the high-symmetry directions near the gamma point, for three different V-I codoping concentrations 2.08\% V and 1.39\% I (Fig. 3b), 4.16\% V and 2.78\% I (Fig. 3c), and 6.25\% V and 4.17\% I (Fig. 3d). For all the codoping concentrations the Fermi level lies inside the gap, which is enlarged from 53 meV (Fig. 3b) to 84 meV (Fig. 3d) when the codoping concentration increases. This proves that the V-I codoping drives the conducting surface into an insulator with an appreciable surface band gap, allowing the formation of high-temperature QAHE.
\begin{figure}
  \includegraphics[width=7cm,angle=0]{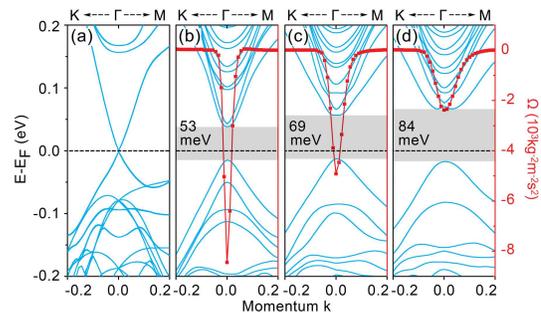}
  \caption{(a) Band structure of a Sb$_2$Te$_3$ thin film calculated in a $4\times4 $ Sb$_2$Te$_3$ supercell with a thickness of 6QLs along high symmetry directions. (b)-(d) Solid lines: Band structures of V-I codoped Sb$_2$Te$_3$ with the film thickness of 6QLs at different codoping concentrations: (b) 2.08\% V and 1.39\% I, (c) 4.16\% V and 2.78\% I, and (d) 6.25\% V and 4.17\% I. Square symbols: Corresponding Berry curvatures.
  } \label{Fig3}
\end{figure}

A separate and unambiguous evidence for the manifestation of QAHE is obtained by integrating the Berry curvature of the occupied valence bands using the expression~\cite{BerryCurvature1,BerryCurvature2}
\begin{equation}
  \Omega (\bm k)=-\sum_n f_n \sum_{n'\neq n } \frac{2{\rm Im} \langle \psi_{n \bm k}|v_x| \psi_{n' \bm k}\rangle \langle \psi_{n' \bm k} |v_y|\psi_{n \bm k}\rangle}{(E_{n'}-E_n)^2},
\end{equation}
where $n$ is the band index, $E_n$ and $ \psi_{n \bm k}$  are the eigenvalue and eigenstate of the band $n$, $v_x=\partial E/\partial k_x$ and $v_y=\partial E/\partial k_y$ are velocity operators along $x$ and $y$ directions within the film plane, and $f_n=1$ for all occupied bands. This integration gives the corresponding Hall conductance~\cite{BerryCurvature2}. The Berry curvature distribution along the high symmetry directions shows a large negative peak near the gamma point and zero elsewhere (Figs. 3(b)-3(d)). As a consequence, the total integration or the Hall conductance with the Fermi level lying inside the band gap must be nonzero. With the increase of the surface band gap, the region with finite Berry curvatures broadens, and the curvature peak gradually decreases, potentially signifying the unchanged Hall conductance. Our quantitative calculations reveal that the Hall conductance is indeed quantized, i.e. $\sigma_{xy}=e^2/h$, for all the three codoping concentrations, and thus that the system is in a QAHE state.

In conclusion, our findings demonstrate that codoping TIs provides a versatile route to realize the QAHE at high temperatures~\cite{notes}. Our proof-of-principle demonstrations revealed that codoping results in: 1) the formation of large band gaps, 2) high Curie temperatures, 3) robust quantized Hall conductance. These manifestations originate from the realization of hard ferromagnetism (in our case due to V-dopants), strong spin-orbit coupling (here because of I-dopants), and, most crucially, the pinning of the Fermi level inside the largely preserved (upon codoping) intrinsic TI bulk gap. Even at very low concentrations, we have shown that the lower-bound temperature for observing the QAHE in our codoped TI is $\sim$ 50 Kelvin, which is three orders of magnitude higher than what was previously observed~\cite{ChangCuiZu,ExperimentalQAHE1,ExperimentalQAHE2}, and by further increasing and tailoring the codoping and TI environment it is feasible to further increase this temperature. So far, the QAHE was experimentally observed by using the host TI (Bi,Sb)$_2$Te$_3$ doped with chromium atoms ~\cite{ChangCuiZu,ExperimentalQAHE1,ExperimentalQAHE2}. These experiments could be readily extended to using the proposed codoping approach to realize high-temperature QAHE. \textit{n}-\textit{p} codoping, while never used in the context of the QAHE, has been previously used, e.g., by ourselves in the studies of different kinds of functional materials~\cite{n-p codoping in TiO2,npCodopedDMS}.

\begin{acknowledgments} 
We are grateful to the financial support of NSFC (51025101, 11104173, 11034006, 11474265, and 61434002) and NBRPC (2014CB921103). Z.Q. acknowledges support from the 100 Talent Program of CAS. S.B.Z. is supported by the US DOE Office of Basic Energy Sciences (DE-SC0002623). H.C. is supported by NSF (DMR-1122603). The supercomputing center of USTC and the National Energy Research Supercomputer Center of the US Department of Energy are gratefully acknowledged for high performance computing assistance.

S.F. Qi and Z.H. Qiao contributed equally to this work.
\end{acknowledgments}


\begin{thebibliography}{99}
\bibitem{AnotherU}
We also found that the increasing of Hubbard U, i.e., $U=6$ eV, gives very limited difference comparing to band structures with $U=3$ eV.
\end{thebibliography}
\end{document}